\documentclass[longauth]{aa} 
\renewcommand{\deg}{^\circ} 

\usepackage[usenames, dvipsnames]{color}
\usepackage{graphicx}

\usepackage{txfonts}

\usepackage[nointegrals]{wasysym}

\usepackage{amssymb}

\usepackage{siunitx}
\DeclareSIUnit\parsec{pc}
\DeclareSIUnit\lightyear{ly}
\DeclareSIUnit\erg{erg}
\DeclareSIUnit\MeV{MeV}
\DeclareSIUnit\GeV{GeV}
\DeclareSIUnit\TeV{TeV}
\DeclareSIUnit\PeV{PeV}

\usepackage{natbib}
\bibpunct{(}{)}{;}{a}{}{,}

\PassOptionsToPackage{american}{babel}
\usepackage{babel}

\PassOptionsToPackage{authoryear}{natbib}
 \usepackage{natbib}

\renewcommand{\deg}{^\circ}

\usepackage[normalem]{ulem}

\begin{document}

   \title{Very-high-energy gamma-ray observations of the Type Ia Supernova SN~2014J with the MAGIC telescopes}

%
\author{
M.~L.~Ahnen\inst{1} \and
S.~Ansoldi\inst{2,}\inst{25} \and
L.~A.~Antonelli\inst{3} \and
P.~Antoranz\inst{4} \and
C.~Arcaro\inst{5} \and
A.~Babic\inst{6} \and
B.~Banerjee\inst{7} \and
P.~Bangale\inst{8} \and
U.~Barres de Almeida\inst{8,}\inst{26} \and
J.~A.~Barrio\inst{9}
% ...
}
\institute { ETH Zurich, CH-8093 Zurich, Switzerland
\and Universit\`a di Udine, and INFN Trieste, I-33100 Udine, Italy
\and INAF National Institute for Astrophysics, I-00136 Rome, Italy
\and Universit\`a  di Siena, and INFN Pisa, I-53100 Siena, Italy
\and Universit\`a di Padova and INFN, I-35131 Padova, Italy
\and Croatian MAGIC Consortium, Rudjer Boskovic Institute, University of Rijeka, University of Split and University of Zagreb, Croatia
\and Saha Institute of Nuclear Physics, 1/AF Bidhannagar, Salt Lake, Sector-1, Kolkata 700064, India
\and Max-Planck-Institut f\"ur Physik, D-80805 M\"unchen, Germany
\and Universidad Complutense, E-28040 Madrid, Spain
\and Inst. de Astrof\'isica de Canarias, E-38200 La Laguna, Tenerife, Spain
\and Universidad de La Laguna, Dpto. Astrof\'isica, E-38206 La Laguna, Tenerife, Spain
\and University of \L\'od\'z, PL-90236 Lodz, Poland
\and Deutsches Elektronen-Synchrotron (DESY), D-15738 Zeuthen, Germany
\and Institut de Fisica d'Altes Energies (IFAE), The Barcelona Institute of Science and Technology, Campus UAB, 08193 Bellaterra (Barcelona), Spain
\and Universit\"at W\"urzburg, D-97074 W\"urzburg, Germany
\and Institute for Space Sciences (CSIC/IEEC), E-08193 Barcelona, Spain
\and Technische Universit\"at Dortmund, D-44221 Dortmund, Germany
\and Finnish MAGIC Consortium, Tuorla Observatory, University of Turku and Astronomy Division, University of Oulu, Finland
\and Unitat de F\'isica de les Radiacions, Departament de F\'isica, and CERES-IEEC, Universitat Aut\`onoma de Barcelona, E-08193 Bellaterra, Spain
\and Universitat de Barcelona, ICC, IEEC-UB, E-08028 Barcelona, Spain
\and Japanese MAGIC Consortium, ICRR, The University of Tokyo, Department of Physics and Hakubi Center, Kyoto University, Tokai University, The University of Tokushima, Japan
\and Inst. for Nucl. Research and Nucl. Energy, BG-1784 Sofia, Bulgaria
\and Universit\`a di Pisa, and INFN Pisa, I-56126 Pisa, Italy
\and ICREA and Institute for Space Sciences (CSIC/IEEC), E-08193 Barcelona, Spain
\and also at the Department of Physics of Kyoto University, Japan
\and now at Centro Brasileiro de Pesquisas F\'isicas (CBPF/MCTI), R. Dr. Xavier Sigaud, 150 - Urca, Rio de Janeiro - RJ, 22290-180, Brazil
% ...
}

   \date{Received ... ; accepted ...}

  \abstract
  {In this work we present data from observations with the MAGIC telescopes of SN~2014J detected in January 21 2014, the closest Type Ia supernova since Imaging Air Cherenkov Telescopes started to operate.}	
   {We probe the possibility of very-high-energy (VHE; $E\geq100$ GeV) gamma rays produced in the early stages of Type Ia supernova explosions.}
   {We performed follow-up observations after this supernova explosion for 5 days, between January 27 and February 2 in 2014. We search for gamma-ray signal in the energy range between 100~GeV and several TeV from the location of SN~2014J using data from a total of $\sim5.5$ hours of observations. Prospects for observing gamma-rays of hadronic origin from SN~2014J in the near future are also being addressed.}
   {No significant excess was detected from the direction of SN~2014J. Upper limits at 95$\%$ confidence level on the integral flux, assuming a power-law spectrum, d$F/$d$E\propto E^{-\Gamma}$, with a spectral index of $\Gamma=2.6$, for energies higher than 300 GeV and 700 GeV, are established at $1.3\times10^{-12}$ and $4.1\times10^{-13}$ photons~cm$^{-2}$s$^{-1}$, respectively.}
   {For the first time, upper limits on the VHE emission of a Type Ia supernova are established. The energy fraction isotropically emitted into TeV gamma rays during the first $\sim10$ days after the supernova explosion for energies greater than 300 GeV is limited to $10^{-6}$ of the total available energy budget ($\sim 10^{51}$ erg). Within the assumed theoretical scenario, the MAGIC upper limits on the VHE emission suggest that SN~2014J will not be detectable in the future by any current or planned generation of Imaging Atmospheric Cherenkov Telescopes.}
  

   \maketitle
%
\section{Introduction}

Type Ia supernovae (SNe) are extremely luminous stellar explosions, which are believed to originate from primary carbon-oxygen white dwarfs (WD) in binary systems reaching the Chandrasekhar mass limit of 1.4 M$_{\odot}$ \citep{Chandrasekhar1931}. When this happens, the electron-degenerate core can no longer support the gravitational pressure, leading to an implosion of the progenitor WD. Thereby the temperature grows up to the carbon fusion point, giving rise to a thermonuclear explosion releasing so much nuclear energy ($\sim10^{51}$~erg; \citealt{1993ApJ...412..192B}) that no compact remnant is expected. The nature of the companion star is still unclear, although two classical progenitor scenarios have been promoted: single-degenerate model, in which the WD accretes material from a red giant star \citep{1973ApJ...186.1007W}, and the double-degenerate model, in which the explosion is produced by the merging of two WDs \citep{1984ApJS...54..335I}. Type Ia SNe have been used to provide information on the Galactic chemical evolution \citep{1995ApJS...98..617T} and to measure cosmological parameters \citep[e.g.][]{Perlmutter:1998np} since they can be used as standard candles thanks to their consistent luminosity \citep{Branch1992}. Still, the evolutionary path that leads to a carbon-oxygen WD which exceeds the Chandrasekhar limit is not well understood yet.\\

SN~2014J was detected on January 21 2014 (MJD 56678) by the UCL Observatory \citep{2014CBET.3792....1F} and classified as a Type Ia SN with the Dual Imaging Spectrograph on the ARC 3.5 m telescope (January 22; \citealt{Goobar2014}). It is located in the starburst galaxy M82 at a distance of 3.6~Mpc \citep{karachentsev_masses_2006}. Its proximity has granted it the title of the nearest Type Ia SN in the past 42 years and motivated large multiwavelength follow-up observations from radio to very-high-energy (VHE; $E\geq100$ GeV) gamma rays. \\
Deep studies of color excess and reddening estimation were carried out on SN 2014J. These studies helped to understand the properties of the dust that affects the brightness of the SN as well as to provide new clues on the progenitors, which are important parameters in the cosmology investigation. \cite{2014ApJ...788L..21A} presented, for the first time, a characterization of the reddening of a Type Ia SN in a full range from $0.2~\mu$m to $2~\mu$m. Their results, with reddening values of $E(B-V)\sim1.3$ and $R_{V}\sim1.4$, are compatible with a power-law extinction, expected in the case of multiple scattering scenarios. In the same wavelength band, from UV to near infrared (NIR), \cite{2014MNRAS.443.2887F} found reddening parameter values of  $E(B-V)\sim1.2$ and $R_{V}\sim1.4$. In this model, the extinction is explained to be caused by a combination of the galaxy dust and a dusty circumstellar medium. However, although compatible with an extinction law with a low value of $R_{V}\sim1.4$ and consistent as well with previous mentioned results, \cite{2015ApJ...805...74B}, making use of \textit{Swift}-UVOT data, suggested that most of the reddening is caused by the interstellar dust. Optical and NIR linear polarimetric observations of the source presented in \cite{2014ApJ...795L...4K} supports the scenario where the extinction is mostly produced by the interstellar dust. These evidences favor the double-degenerate scenario for SN~2014J, where less circumstellar dust is expected than in cases with a giant companion star. This type of companion are indeed ruled out by several authors as possible progenitors in SN~2014J, e.g. \cite{pereztorres2014}, with the most sensitive study in the radio band of a Type Ia SN, or \cite{Margutti:2014uha} in the X-ray band. The former reported non-detection from the observations performed with eMERLIN and EVN. These results, compared with a detailed modeling of the radio emission from the source, allowed to exclude the single-degenerate scenario in favour of the double-degenerate one with constant density medium of $n \lesssim1.3$ cm$^{-3}$. 

Several authors have speculated about the possibility of SN explosions being able to produce gamma-ray emission at detectable level by current and/or future telescopes. However, these models generally consider Type II SNe due to the strong wind of the progenitors (e.g. \citealt{1995A&A...293L..37K} and \citealt{2009A&A...499..191T}). Nevertheless, given the proximity of SN~2014J, this event provides a good exploratory opportunity to probe the eventual production of VHE gamma rays during the first days after such an explosion. 

In this work, we present the analysis results of SN~2014J observations performed with the MAGIC telescopes.


\section{Observations $\&$ Results}

\begin{table*}[ht]\footnotesize
	\centering
	\begin{tabular}{ c  c c  c  c c}
		\hline
		\hline
		 \multicolumn{2}{c}{Date} & Eff. Time & Zd & UL ($E>300$ GeV) & UL ($E>700$ GeV)\\
		 \cline{1-2}
		[yyyy-mm-dd] & [MJD] & [h] & [degrees] & [photons~cm$^{-2}$ s$^{-1}$] & [photons~cm$^{-2}$ s$^{-1}$]\\
		\hline
		2014-01-27 & 56684.23 & 0.43 & 47-50 & $1.03\times10^{-11}$ &$-$ \\
		2014-01-28 & 56685.06 & 1.41 & 40-43 & $2.19\times10^{-12}$ & $1.55\times10^{-12}$\\
		2014-01-29 & 56686.09 & 1.30 & 40-42 & $4.55\times10^{-12}$ & $5.97\times10^{-13}$\\
		2014-02-01 & 56689.07 & 0.98 & 40-42 & $3.14\times10^{-12}$ & $9.98\times10^{-13}$\\
		2014-02-02 & 56690.08 & 1.30 & 40-42 & $3.35\times10^{-12}$ & $1.76\times10^{-12}$ \\
		Total & $-$ & 5.41 & 40-50 &  $1.30\times10^{-12}$ &  $4.10\times10^{-13}$\\ 
		\hline
	\end{tabular}
	\caption{Summary of the MAGIC observations of SN~2014J. From left to right: date of the beginning of the observations, also in Modified Julian Date (MJD), effective time, zenith angle range and integral ULs at 95\% CL above 300 and 700 GeV. The last row reports the integral ULs derived with the entire data sample. Due to low statistics, no integral UL was computed for energies above 700 GeV for the first day of observations.}
	\label{tab:dailyuls}
\end{table*}

\begin{figure}
		\centering
		\includegraphics[width=\hsize]{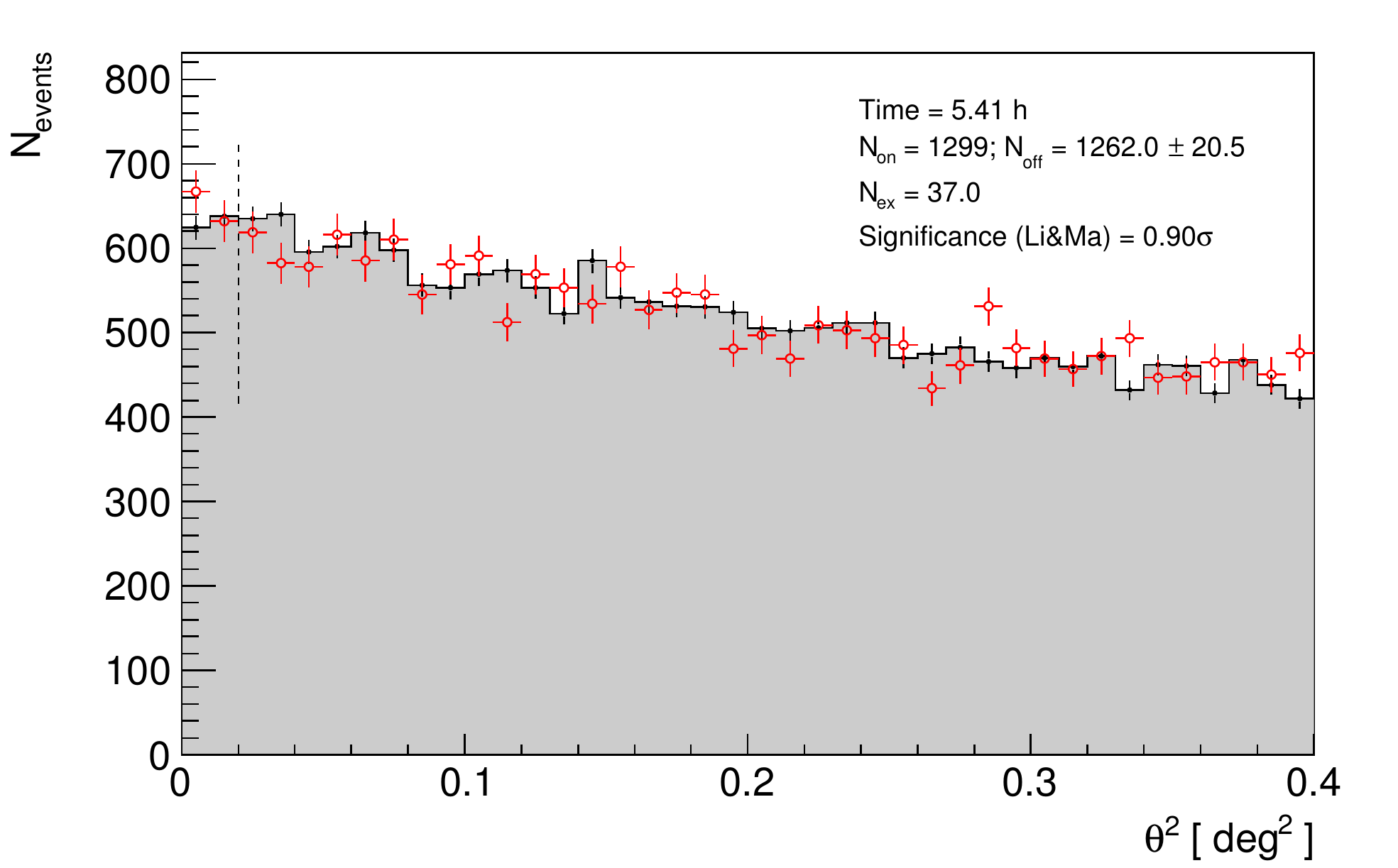}
		\caption{Distribution of the squared angular distance, $\theta^{2}$, after 5.41 hours of observation between the reconstructed arrival direction of the gamma-ray candidate events and the position of the source in the camera (red empty circles). The $\theta^{2}$ distribution of the background events (black points) is also displayed. The vertical dashed line at $\theta^{2}$=0.02 deg$^{2}$ defines the expected signal region.}
		\label{fig:theta2LE}
\end{figure}

\begin{figure}
		\centering
		\includegraphics[width=\hsize]{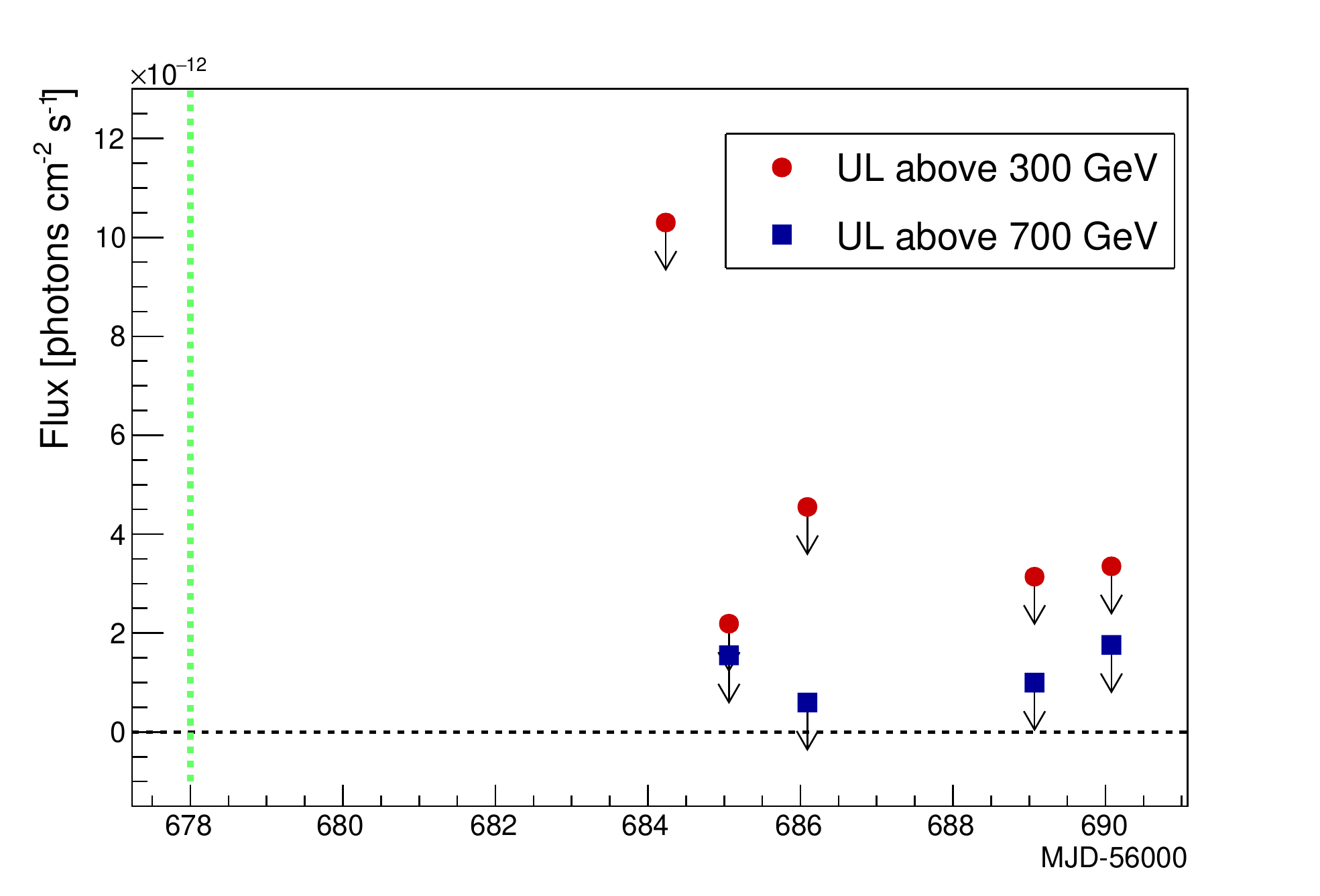}
		\caption{MAGIC daily integral ULs from the direction of SN~2014J for energies above 300 GeV (red circles) and 700 GeV (blue squares). The integral UL for energies above 700 GeV was not computed for the first night (MJD 56684) due to low statistics (see also Table \ref{tab:dailyuls}). The horizontal black dashed line indicates zero flux level and the vertical green line indicates the day of the SN explosion (MJD 56678), just six days before the beginning of the MAGIC observations.}
		\label{fig:dailyuls}
\end{figure}

The MAGIC stereo system at the Observatorio del Roque de los Muchachos on the Canary island of La Palma, Spain (28.8$\deg$N, 17.8$\deg$ W, 2200 m a.s.l.), consists of two 17 m Imaging Air Cherenkov Telescopes (IACTs). The MAGIC telescopes reach one of the lowest trigger energy thresholds among current IACTs (50 GeV). The observations were carried out in stereoscopic mode, which means that only shower images seen in both telescopes are recorded and analyzed. This mode provides a sensitivity of 0.66 $\pm$ 0.03 \% of the Crab Nebula flux in 50 hours of observation for energies above 220 GeV \citep{2016APh....72...76A}. \\

SN~2014J was observed under moderate moonlight conditions from January 27 to 29 and on February 1 and 2 under dark-night conditions at medium zenith angles (from 40$\deg$ to 52$\deg$). The MAGIC observations started six days after the first detection by the UCL Observatory because of the adverse weather conditions. The complete data set up to 50$\deg$ ($\sim5.5$ hours) was used for the analysis given the overall good quality of the data (concerning weather, light conditions and performance of the system).\\

Figure \ref{fig:theta2LE} shows the distribution of squared angular distance ($\theta^2$) between the reconstructed gamma-ray direction and the position of either SN~2014J (on-source histogram) or the center of the background control region (off-source histogram). The resulting excess of the on-source histogram over the background from the region, where gamma-ray events from SN~2014J are expected is compatible with zero excess. The significance computed using Eq. 17 of \cite{LiMa1983} is $0.90 \sigma$.

Upper limits (ULs) on the flux were computed following the \citet{Rolke2005} method for 95\% confidence level (CL), assuming a Gaussian background and a systematic uncertainty of 30\% on the effective area of the instrument \citep{Albert2008}. The spectrum that we assumed was a power-law function, d$F/$d$E \propto E^{-\Gamma}$, with a spectral index of 2.6. Variations of $\sim$20\% in the spectral index produced changes in the integral ULs of less than 5\%, and hence small deviations from the used spectral index do not critically affect the reported ULs. The ULs above 300 GeV and 700 GeV for the single-night observations are reported in Table \ref{tab:dailyuls} and depicted in Figure \ref{fig:dailyuls}. 

After $\sim5.5$ hours of observations with the MAGIC telescopes, we establish an integral UL on the gamma-ray flux for energies above 300 GeV of $1.3\times10^{-12}$ photons~cm$^{-2}$s$^{-1}$ at 95\% CL, which corresponds to $\sim 1.0 \%$ in units of the Crab Nebula flux (CU) in the same energy range. For energies above 700 GeV, the integral UL is $4.1\times10^{-13}$ photons~cm$^{-2}$s$^{-1}$, corresponding to $\sim 1.1 \%$ CU at the same CL. Our ULs for $E>700$ GeV are already close to the flux from the host galaxy M82 measured by VERITAS in the same energy range, $(3.7\pm0.8_{stat}\pm0.7_{syst})\times10^{-13}$ photons~cm$^{-2}$ s$^{-1}$ \citep{2009Natur.462..770V}, which constitutes an irreducible background for our measurement. Under the hypothesis that M82 has a gamma-ray spectrum of d$F/$d$E = 3\times 10^{-16} (E/1000$ GeV$)^{-2.5}$ photons~cm$^{-2}$ s$^{-1}$ GeV$^{-1}$, as measured by VERITAS, the expected number of excess events in our observations would be 9.4, with a 95\% CL lower limit at -6.1. The observed number of excess events by MAGIC is -4.2, with an associated p-value of $8.4\times10^{-2}$, hence consistent with the VHE flux of M82 measured by VERITAS. 

\section{Discussion}
VHE gamma rays are typically secondary products of particle acceleration to $\gtrsim$~TeV energies, either in hadronic or leptonic processes \citep{2013APh....43...71A}. The primary particles of such processes can either be protons or electrons. In the first case, the gamma-ray emission is produced by neutral pions from the inelastic collisions between the protons accelerated in the SN and the ambient atomic nuclei. In the latter case, the gamma rays result from inverse Compton (IC) process of the accelerated electrons on the ambient photons. In both cases, the environment plays an important role for the production of VHE gamma-ray radiation. A near and young supernova ($\sim1$~week old) emitting in this energy regime could shed light on the progenitors of these thermonuclear stellar explosions.\\

Although we did not detect VHE gamma rays right after the explosion, using the known distance of M82 \citep[$d_\mathrm{M82} = 3.6$~Mpc][]{karachentsev_masses_2006} and assuming, as before, a spectral index of 2.6, one can convert the measured flux UL into an UL on the power emitted into VHE gamma rays. Therefore, given the integral UL for energies greater than 300 GeV, $1.3\times10^{-12}$photons~cm$^{-2}$s$^{-1}$, the resulting UL on the power emitted is of the order of $10^{39}$ erg s$^{-1}$. If one now assumes an emission period of the order of 10 days, the total energy emitted in VHE gamma rays during this period is smaller than $10^{45}$ erg, which is about $10^{-6}$ of the total available energy budget of the SN explosion.\\

Models of the evolution of young supernova remnants (SNRs) can be used to estimate the expected emission from the region in the future. One of the most important parameters to be assumed is the density profile of the SN ejecta. In this work, we considered a simple power-law density profile, which allows us to use the \cite{Dwarkadas:2013nwa} model to obtain an analytic solution for the estimated flux. Other density profiles have been used in the literature: Models like W7 or WDD1, applied by \cite{Nomoto} and \cite{1999ApJS..125..439I}, are usually utilized in Type Ia SN studies, but they are based on the single-degenerate scenario. \cite{1998ApJ...497..807D} discussed a possible exponential density profile, which could represent better the SN ejecta structure than the power-law one. However, the exponential profile cannot provide an analytic result as the one assumed in this work, which can give a correct solution within the order of magnitude, as explained in  \cite{Dwarkadas:2013nwa}. \\
Thus, making use of  Eq.~10 in \cite{Dwarkadas:2013nwa}, one can obtain the time-dependent emission assuming a hadronic origin. Considering only this hadronic origin, we can establish a lower limit on the total gamma-ray radiation. As discussed above, gamma-ray emission can be also expected from IC processes. Nevertheless, purely leptonic scenarios have been studied and discarded by several authors, e.g. \cite{2008A&A...490..515V}.\\

The expected flux depends strongly not only on the assumed density structure of the SNR but also on the density profile of the surrounding interstellar medium (ISM).  As shown by different authors, we can consider double-degenerate scenario in the case of SN 2014J, i.e. two WDs progenitors. WDs do not suffer wind-driven mass-loss and therefore, they are not expected to modify the surrounding medium. Nevertheless, different assumptions, from the lack of certainty on the progenitors, have also been studied (see e.g.  \citealt{2000ApJ...541..418D}). We can then assume that the Type Ia SN explosion took place in a uniform density medium. In this work, we used a density of $n=2.2\times10^{-24}$ g/cm$^{3}$, based on \cite{pereztorres2014}, assuming that all the content in the host galaxy of our source, M82, stems from the neutral hydrogen, H$_{\text{I}}$. This homogeneous medium assumption leads to an increasing flux emission, above a certain gamma-ray energy, with time in the free-expansion SNR stage, as shown below in the expression given  by \cite{Dwarkadas:2013nwa}:

\begin{equation}
\label{eq:Dwarkadasflux}
		F_{\gamma}(>1 \text{TeV},t)=\frac{3q_{\gamma}\xi(\kappa C_{1})^{5}m^{3}}{6(5m-2)\beta\mu m_{p}d^{2}}n^{2}t^{5m-2}
\end{equation}

where the assumed parameters in this work are

\begin{itemize}
\item $q_{\gamma} = 1\times10^{-19}$~cm$^{3}$s$^{-1}$erg$^{-1}$H-atom$^{-1}$ (for energies greater than 1 TeV) is the emissivity of gamma rays normalised to the cosmic ray energy density tabulated in \cite{Drury1994}. This value corresponds to a spectral index of 4.6 of the parent cosmic ray distribution, which was selected according to the assumed spectral index in this work, $\Gamma=2.6$
\item $\xi = 0.1$ is the fraction of the total SN explosion energy converted to cosmic ray energy, so an efficient cosmic ray acceleration is assumed
\item $\kappa = 1.2$ is the ratio between the radius of the forward shock and the contact discontinuity (which separates ejecta and reverse shock)
\item $C_{1} = 1.25\times10^{13}$cm/s$^{m}$ is referred to as a constant related to the kinematics of the SN. This value is calculated from the relation given by \cite{Dwarkadas:2013nwa},  $R_{shock}=\kappa C_{1}t^{m}$. In turn, $R_{shock}$ is obtained from Eq.~2 in \cite{Gabici2016}, by assuming an explosion energy of $10^{51}$ erg, a mass of the ejecta of $1.4$ M$_{\odot}$ and a ISM density of $1.3$ cm$^{-3}$, whose value is constrained by \cite{pereztorres2014}
\item $\beta = 0.5$ represents the volume fraction of the already shocked region from which the emission arises 
\item $\mu = 1.4$ is the mean molecular weight
\item $m_{p} = 1.6\times10^{-24}$g is the proton mass 
\item $d = 3.6$ Mpc is the distance to our source
\item $t$ is the elapsed time since the explosion, and 
\item $m$ is the expansion parameter. 
\end{itemize}

The expansion parameter varies along the free-expansion phase in different ways according to the assumed model for the density structure of the SN ejecta after the explosion. In this work, we make use of the power-law profile with a density profile proportional to $R^{-7}$ \citep{1982ApJ...258..790C}, where $R$ is the outer radius of the ejecta. The initial value of the expansion parameter is very unalike depending on the density profile assumed, but in all cases evolve to $m=0.40$ \citep{1998ApJ...497..807D}. This limit at 0.40 is constrained by the beginning of the Sedov-Taylor phase. \\
The expansion parameter for the power-law profile keeps constant at 0.57 in the first years of the free-expansion stage. Given this value, the expected flux above 1 TeV  (constrained by the emissivity of gamma rays, $q_{\gamma}$, tabulated in \citealt{Drury1994}) at the time of the MAGIC observations ($t=6$ days) from Eq. \ref{eq:Dwarkadasflux} is $\approx 10^{-24}$ photons~cm$^{-2}$s$^{-1}$. This flux is consistent with the UL at 95\% CL derived from MAGIC data in the same energy range, $2.8\times10^{-13}$ photons~cm$^{-2}$s$^{-1}$ and hence, the power-law density profile could be considered as possible model to describe the density structure of SN~2014J, considering all the assumptions and parameters selection discussed above. On the other hand, this model predicts a constant parameter of $m=0.57$ during the first $\sim300$ years, after which it starts dropping gradually \citep{1998ApJ...497..807D}. Although the flux keeps increasing with time according to Eq. \ref{eq:Dwarkadasflux}, with this low expansion parameter it will still be about $10^{-21}$ photons~cm$^{-2}$s$^{-1}$ 100 years after the SN occurred, which is well below the sensitivity of the current and planned VHE observatories (several orders of magnitude below the sensitivity that the future Cherenkov Telescope Array, CTA\footnote{as shown in https://www.cta-observatory.org/science/cta-performance/}, will reach).
\section{Conclusions}

The MAGIC telescopes performed observations of the nearest Type Ia SN in the last decades, SN\,2014J. No VHE gamma-ray emission was detected. Integral ULs for energies above 300 GeV and 700 GeV were established at $1.3\times10^{-12}$ photons~cm$^{-2}$s$^{-1}$ and $4.1\times10^{-13}$ photons~cm$^{-2}$s$^{-1}$, respectively, for a 95\% CL and assuming a power-law spectrum. The flux UL at $E>\SI{300}{\GeV}$ corresponds to an emission power of $< 10^{39}$~erg~s$^{-1}$ or a total maximal emitted VHE gamma-ray energy during the observational period -- approximately ten days -- of $< 10^{45}$ erg, which is about $10^{-6}$ times the total energy budget of a Type Ia SN explosion ($\sim 10^{51}$ erg). Following \cite{Dwarkadas:2013nwa} model for hadronic gamma-ray flux, a power-law density profile proportional to $R^{-7}$ is consistent with our ULs, although, due to the uncertainties in several parameters, this cannot exclude other, more sophisticated, theoretical scenarios. Assuming this SN density profile and a constant density medium, we can estimate an expected emission from the region of the source of $\approx 10^{-24}$ photons~cm$^{-2}$s$^{-1}$. Following these assumptions, this flux would not increase enough in a near future to be detectable by any current or future generation of IACTs, as CTA.

\begin{acknowledgements}
We would like to thank
the Instituto de Astrof\'{\i}sica de Canarias
for the excellent working conditions
at the Observatorio del Roque de los Muchachos in La Palma.
The financial support of the German BMBF and MPG,
the Italian INFN and INAF,
the Swiss National Fund SNF,
the ERDF under the Spanish MINECO
(FPA2015-69818-P, FPA2012-36668, FPA2015-68278-P,
FPA2015-69210-C6-2-R, FPA2015-69210-C6-4-R,
FPA2015-69210-C6-6-R, AYA2013-47447-C3-1-P,
AYA2015-71042-P, ESP2015-71662-C2-2-P, CSD2009-00064),
and the Japanese JSPS and MEXT
is gratefully acknowledged.
This work was also supported
by the Spanish Centro de Excelencia ``Severo Ochoa''
SEV-2012-0234 and SEV-2015-0548,
and Unidad de Excelencia ``Mar\'{\i}a de Maeztu'' MDM-2014-0369,
by grant 268740 of the Academy of Finland,
by the Croatian Science Foundation (HrZZ) Project 09/176
and the University of Rijeka Project 13.12.1.3.02,
by the DFG Collaborative Research Centers SFB823/C4 and SFB876/C3,
and by the Polish MNiSzW grant 745/N-HESS-MAGIC/2010/0.
\end{acknowledgements}


\bibliographystyle{aa}
\bibliography{bibliography}

\end{document}